\documentclass[superscriptaddress,aps,pra,twocolumn,showpacs,floatfix]{revtex4-2}

\usepackage[utf8]{inputenc}
\usepackage{graphicx,amsmath,amsfonts,amssymb}
\usepackage{xcolor}
\usepackage[colorlinks=true,allcolors=blue]{hyperref}
\usepackage{float}
\usepackage{placeins}
\usepackage{subcaption}
\usepackage{lipsum} 
\newcommand{\orcid}[1]{\href{https://orcid.org/#1}{\includegraphics[width=7pt]{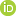}}}

\begin{document}

\preprint{APS/123-QED}

\title{Probing non-adiabaticity in a quantum Otto cycle through quantum texture and imaginarity}

\author{Jonas F. G. Santos\orcid{0000-0001-9377-6526}}
\email{jonassantos@ufgd.edu.br}
\affiliation{Faculdade de Ci\^{e}ncias Exatas e Tecnologia, Universidade Federal da Grande Dourados, Caixa Postal 364, 79804-970, Dourados, MS, Brazil}

\begin{abstract}
Finite-time driving in quantum thermodynamic cycles inevitably generates coherences in the instantaneous energy basis, leading to quantum friction and degrading the operational performance. In this work, we demonstrate that the non-adiabatic inner friction produced during unitary driving of a two-level system can be faithfully quantified by combining two distinct resource-theoretic notions: quantum texture (rugosity) and imaginarity. We show that neither rugosity nor imaginarity alone is sufficient to track the total non-adiabaticity, and we introduce a combined witness, which is strictly non-negative and vanishes if and only if the driving is quasi-static. Applying this framework to a finite-time quantum Otto refrigerator, we establish an exact closed-form relation connecting the coefficient of performance directly to the quantum texture and imaginarity, evaluated at the end of the unitary compression and expansion strokes. Furthermore, we show that local minima of the non-adiabatic witness accurately identify the operating regimes that maximize cooling performance. Our findings provide a geometric and resource-theoretic foundation for inner friction, offering practical diagnostic tools for optimizing finite-time quantum thermal machines.
\end{abstract}

\maketitle

\section{Introduction}

Recent advances in quantum science and technology have shown that a precise understanding of how to manipulate and preserve quantum features, such as quantum coherence and entanglement, is crucial to effectively mitigating dissipation and errors, as well as to enhancing the performance of various quantum devices, ranging from thermodynamic cycles \cite{PhysRevLett.123.240601,Kosloff2017-bt,PhysRevA.99.062103, PRXQuantum.3.040323, Lostaglio2015-mb, Hammam2022-md, PhysRevE.102.062152, PhysRevE.97.062105, PhysRevA.107.012221} to metrological protocols ~\cite{Giovannetti2011-ff, Huang2024-th, nkm3-qply,PhysRevA.111.052404}, including quantum computation and related areas ~\cite{StreltsovRMP2017, NielsenChuang2010, ChitambarRMP2019}. In these different scenarios, finite-time dynamics is crucial to approaching idealized developments from realistic devices. Whenever a finite-time driving is accompanied by a Hamiltonian that does not commutes at different times, coherence in the energy basis is generated ~\cite{PhysRevE.73.025107, PhysRevLett.113.260601, PhysRevA.99.062103, PhysRevLett.123.240601}, with profound impacts on the performance on quantum protocols.

In quantum metrology protocols, quantum coherence has been widely recognized as a valuable resource that enhances the precision of parameter estimation~\cite{Giovannetti2011-rn, PhysRevA.111.052404}. Conversely, in finite-time quantum thermodynamics, the buildup of coherence in the instantaneous energy basis is typically associated with inner friction, ultimately degrading the performance of quantum thermal devices~\cite{Kosloff2017-bt,PhysRevLett.123.240601, PhysRevB.86.014501, PhysRevA.99.062103, hpkj-srj6}. Nevertheless, actively managing coherences across consecutive strokes—for instance, through incomplete thermalization with heat sources—has emerged as a viable route to mitigate irreversible losses at short operational times~\cite{PhysRevA.99.062103, Sutantyo2024-lu}. A standard method to assess this non-adiabaticity relies on the quantum relative entropy, where the instantaneous state is compared to a quasi-static reference state that remains diagonal in the instantaneous energy basis~\cite{Santos2019-lf, PhysRevLett.113.140401}. Although this entropic framework is remarkably powerful, it lacks direct resolution of the geometric features and internal phase structures of quantum coherences. For a single qubit, such geometric insight can be comprehensively accessed through the trajectory and constraints of the state within the Bloch sphere.

In the last years, two quantum resources have emerged within a geometric connection with coherence of a quantum state for a given basis. The first is the quantum texture \cite{PhysRevLett.133.260801, Chen2026-ph, Vieira2026-qn}. It quantifies how disparate a quantum state is from a completely flat reference state for a given basis. Interestingly, quantum texture is non-zero even for thermal states, due to the imbalance between the populations. Quantum texture has been used in diverse applications, such as characterization of quantum gates ~\cite{PhysRevLett.133.260801, Vieira2026-qn}, its connection with quantum speed limits ~\cite{Xie2026-kw} and in other interesting scenarios ~\cite{hts6-k7d3, 83pf-837s, y2wz-ypln, Muthuganesan2026-as, Cao2026-zj, Wang2026-ag}. The second resource is the imaginarity \cite{Hickey2018-as, PhysRevLett.126.090401, kj4y-ttqb}, in which the resource arises from imaginary terms that appear in the density matrix of a quantum state for a given basis. Here the reference states are those with only real elements in their density matrices. Different advances have been made in extending ways of quantifying imaginarity of quantum states ~\cite{s7kr-8rrn, PhysRevA.108.062203}, with many applications in quantum information theory ~\cite{kj4y-ttqb, PhysRevLett.133.190201, PhysRevA.103.032401, PhysRevA.110.052202, np2q-68s8}. Despite their clear connections to quantum coherence, the study of quantum texture and imaginarity, particularly their complementary role in quantum thermodynamics, remains largely unexplored.

In this work, we introduce a faithful non-adiabaticity witness based on the combination of quantum texture (rugosity) and imaginarity, capable of tracking coherence generation in general unitary protocols as well as in quantum Otto cycles fueled by a single qubit. Evaluated with respect to the instantaneous basis of the driven Hamiltonian, this quantifier is strictly non-negative and vanishes if and only if the state remains diagonal throughout the dynamics. Furthermore, we establish an analytical relation between the performance of finite-time Otto cycles and this witness, demonstrating that it directly quantifies the inner friction generated across the driven strokes

The manuscript is organized as follows. In Sec.~\ref{Quantum texture and imaginarity}, we introduce the concepts of quantum texture (rugosity) and imaginarity defined with respect to the instantaneous basis of a driven Hamiltonian. We then construct the non-adiabaticity witness for a single qubit and express the inner friction generated in arbitrary unitary dynamics as an explicit function of these geometric resources. Section~\ref{Non-adiabatic witness in a Quantum Otto cycle} is devoted to evaluating the non-adiabaticity witness in a finite-time quantum Otto cycle fueled by a single qubit as the working substance, focusing on the refrigeration regime where both the coefficient of performance and the inner friction are formulated in terms of texture and imaginarity. Finally, concluding remarks and future perspectives are presented in Sec.~\ref{Conclusion}.

\section{Quantum texture and imaginarity}\label{Quantum texture and imaginarity}

Let us consider a general unitary process described by a time-dependent Hamiltonian $H(t)$ in a finite-time $t\in \left[0,\tau\right]$. Denoting by $E_n^t$ and $|E_n^t\rangle$ the instantaneous eigenvalues and eigenstates of the Hamiltonian, respective,  one has
\begin{equation}
    H(t)|E_n^t\rangle  = E_n^t|E_n^t\rangle.
\label{01}
\end{equation}

Suppose a single qubit in an initial state $\rho(0)$ that commutes with the Hamiltonian at time $t=0$. For any time $t$ the state can be written in the instantaneous basis of the Hamiltonian as
\begin{equation}
    \rho(t) = \sum_{n,m = 0}^1\rho_{n,m}(t)\,|E_n^t\rangle \langle E_m^t|.
    \label{02}
\end{equation}

By writing the decomposition in the Bloch sphere for this instantaneous density matrix, we have
\begin{eqnarray}
\rho(t)&=&\frac{1}{2}\Big(\mathbb{I}+\vec{r}(t)\cdot \vec{\sigma}\Big)\nonumber\\
&=&\frac{1}{2}\Big(\mathbb{I}+r_x(t)\,\sigma_x+r_y(t)\,\sigma_y+r_z(t)\,\sigma_z\Big),
\label{Bloch}
\end{eqnarray}
where $\vec{r}(t)=(r_x(t),r_y(t),r_z(t))$ is the Bloch vector and $\vec{\sigma}=(\sigma_x,\sigma_y,\sigma_z)$ are the Pauli matrices. In terms of the matrix elements in Eq.~(\ref{02}),
\begin{equation}
\begin{aligned}
 r_x(t) &= \rho_{01}(t)+\rho_{10}(t)=2\,\mathrm{Re}\,\left[\rho_{01}(t)\right],\\
 r_y(t) &= i\big(\rho_{10}(t)-\rho_{01}(t)\big)=-2\,\mathrm{Im}\,\left[\rho_{01}(t)\right],\\
 r_z(t) &= \rho_{00}(t)-\rho_{11}(t),
\end{aligned}
\label{03}
\end{equation}
with $\rho_{00}(t)+\rho_{11}(t)=1$.

\textit{Instantaneous quantum texture}. The quantum texture has been introduced in Ref. ~\cite{PhysRevLett.133.260801} where, for a fixed basis $\lbrace|i\rangle \rbrace$, there is only a single quantum state in which all matrix elements have the same value,
\begin{equation}
    |f_1\rangle = \frac{1}{\sqrt{D}}|i\rangle 
\end{equation}
where $D$ is the dimension of the associated Hilbert space $\mathcal{H}$. As introduced in Ref. ~\cite{PhysRevLett.133.260801}, the expression $\langle f_1|\rho|f_1\rangle$ serves as a quantifier of the texture of $\rho$. Based on it, the rugosity of the state $\rho$ is defined as $\mathcal{R} = - \ln \langle f_1|\rho|f_1\rangle$.

For a qubit in the instantaneous basis of $H(t)$, we identify the instantaneous textureless state as 
\begin{equation}
    |\tilde{+}\rangle = \frac{1}{\sqrt{2}}\left(|E^t_0\rangle + |E^t_1\rangle \right),
\end{equation} 
and the rugosity of a time-dependent state $\rho(t)$ is given by

\begin{eqnarray}
\mathcal{R}(t) &=& - \ln \langle \tilde{+}|\rho|\tilde{+}\rangle\nonumber\\
     &=& \ln\left[\frac{2}{1+ r_x(t)}\right].
     \label{A03}
\end{eqnarray}

From Eq. (\ref{A03}) we immediately write the instantaneous Bloch component in the $x$-direction as $r_x(t) = 2e^{-\mathcal{R}(t)}-1$.

For quasi-static driving in which the state remains diagonal in the instantaneous basis (no coherences), one has $r_x(t)=0$ and the rugosity reduces to $\ln 2$.

\textit{Instantaneous imaginarity}. Analogous to coherence and quantum texture, imaginarity is a basis dependent quantity. Again, for a fixed basis $\lbrace|i\rangle \rbrace$ for the Hilbert space $\mathcal{H}$, the reference states (those free of imaginarity or reals) are defined as 
\begin{equation}
    \sigma = \sum_{ij} \sigma_{ij} |i\rangle \langle j|,
\end{equation}
where each $\sigma_{ij} \in \mathbb{R}$. In Ref. ~\cite{Hickey2018-as}, they show that, denoting by $F$ the set of all real states $\sigma$ and given a state $\rho$, the following quantity 
\begin{eqnarray}
M(\rho) &:=&\text{min}_{\sigma \in F} ||\rho - \sigma||_1\nonumber\\
     &=& \frac{1}{2}||\rho - \rho^T||_1,
\end{eqnarray}
with $|| \cdot||_1$ the trace norm, is a faithful measure of imaginarity of $\rho$. 

For a qubit in the instantaneous basis of a given time-dependent Hamiltonian $H(t)$, and with components of the Bloch sphere given by Eq. (\ref{03}), the imaginarity is given by
\begin{eqnarray}
  M\left(t\right) &=& \frac{1}{2}||r_y(t) \sigma_y||_1\nonumber\\
  &=& |r_y(t)|.
  \label{A02}
\end{eqnarray}

In the quasi-static regime, the instantaneous component $r_y(t)$ vanishes and the imaginarity is zero.

For the reminder of the work, we refer to rugosity and imaginarity always with respect to the instantaneous basis of a time-dependent Hamiltonian $H(t)$. 

In the following, we want to use the notion of rugosity and imaginarity to account the non-adiabaticity of a given protocol. Firstly, note from Eq. (\ref{03}) that in order to account in a correct manner the coherences generated in the energy basis, we need to use simultaneously the rugosity and the imaginarity, encompassing possible coherence phases . In addition, for any qubit, the inequality $r_x^2(t) + r_y^2(t)+ r_z^2(t) \leq 1$ holds, with the identity only for pure states. By replacing $r_x(t) = 2e^{-\mathcal{R}(t)}-1$ and $M(t) = |r_y\left(t\right)|$ we immediately obtain
\begin{eqnarray}
    M^2\left(t\right) \leq 4e^{-\mathcal{R}\left(t\right)}\left(1 - e^{-\mathcal{R}\left(t\right)}\right) - r_z^2\left(t\right).
    \label{Teo01}
\end{eqnarray}
Furthermore, the combination of the rugosity and imaginarity fully accounts the amount of coherences in the energy basis, i.e.,
\begin{equation}
    \left(2e^{-\mathcal{R}(t)}-1\right)^2 + M^2\left(t\right) = r_x^2(t) + r_y^2(t)  = 4|\rho_{01}|^2.
    \label{cc}
\end{equation}

Equations (\ref{Teo01}) and (\ref{cc}) motivate the following witness for non-adiabaticity of a qubit in a unitary process described by a time-dependent Hamiltonian $H(t)$: consider a qubit initially prepared in a Gibbs state that commutes with the Hamiltonian at $t=0$. For any finite-time process with time duration $\tau$, the quantity 
\begin{equation}
    \mathcal{C}_{\text{diab}}(\tau) \equiv \left(2e^{-\mathcal{R}(\tau)}-1\right)^2  + M^2\left(\tau\right),
    \label{ccc}
\end{equation}
with $\mathcal{R}(\tau)$ and $M\left(\tau\right)$ computed in the instantaneous basis of $H(\tau)$, serves as a witness to account non-adiabaticity of the process. This definition ensures that from the quantum adiabatic theorem, for $\tau$ long enough, the qubit ends the protocol at the state $\rho(\tau) = \sum_n \rho_{n}^\tau|E_n^\tau\rangle \langle E_n^\tau|$, leading to 
$\mathcal{R}^{\text{ad}}(\tau) = \ln(2)$ and $M^{\text{ad}}\left(\tau\right) = 0$, and hence $\mathcal{C}^{\text{ad}}_{\text{diab}}(\tau) = 0$. Otherwise, for any finite-time process the qubit states will not be diagonal in the instantaneous basis, leading to $|\rho_{01}^\tau|>0$ and then $\mathcal{C}_{\text{diab}}(\tau)  = 4|\rho_{01}^\tau|^2 > 0$. Thus, $\mathcal{C}_{\text{diab}}(\tau)$ is a non-negative quantity that is zero if and only if the process is completely quasi-static. 

An essential feature of the proposed witness is its monotonicity decreasing under incoherent dynamics. For a single qubit, the non-adiabaticity witness connects directly to the $\ell_1$-norm of  coherence as $\mathcal{C}_{\text{diab}}(\rho) = [C_{\ell_1}(\rho)]^2$, with  $C_{\ell_1}(\rho) = \sum_{i \neq j} |\rho_{ij}| = 2|\rho_{01}|$~\cite{PhysRevLett.113.140401}.  Because $C_{\ell_1}$ is a \textit{bona-fide} coherence monotone under any incoherent completely positive  and trace-preserving (ICPTP) map $\Phi$, and given that $f(x) = x^2$ is monotonically increasing 
for $x \ge 0$, it immediately follows that
\begin{equation}
\mathcal{C}_{\text{diab}}(\Phi(\rho)) \le \mathcal{C}_{\text{diab}}(\rho).
\end{equation}
This property guarantees that $\mathcal{C}_{\text{diab}}$ monotonically decreases toward zero 
during the uncoupled isochoric thermalization strokes governed by Lindblad dynamics in the energy basis, as we will explore in the next section.

We can use the quantity $\mathcal{C}_{\text{diab}}(\tau)$ to compute the work associated with the inner friction in a unitary evolution where the Hamiltonian does not commute in different times , defined as \cite{PhysRevLett.113.260601}
\begin{equation}
    \langle W_{\text{fric}}(\tau)\rangle  = \langle W (\tau)\rangle  - \langle W(\tau)\rangle_{qs},
    \label{fric}
\end{equation}
where the first and second term on the right side represents the actual work performed on the system and the ideal one, respectively. Considering a unitary dynamics driven by a time-dependent Hamiltonian $H(t)$ followed by a complete thermalization with a Markovian thermal bath at inverse temperature $\beta_0$, in Appendix \ref{A} we derive the following relation
\begin{equation}
    \langle W_{\text{fric}}(\tau)\rangle  = \frac{\hbar \omega(\tau)}{2}\frac{\mathcal{C}_{\text{diab}}(\tau)}{\Delta p_0 + \sqrt{\Delta p_0^2 - \mathcal{C}_{\text{diab}}(\tau)}},
    \label{res01}
\end{equation}
where $\Delta p_0 \equiv \tanh \left(\beta_0 \hbar \omega(0)/2\right)$, as well as the inequality 
\begin{equation}
    \langle W_{\text{fric}}(\tau)\rangle \leq \frac{\hbar \omega(\tau)}{2} \sqrt{\mathcal{C}_{\text{diab}}(\tau)},
    \label{res02}
\end{equation}

The inequality in Eq.~(\ref{res02}) provides a universal geometric upper bound on the internal friction, holding for any unitary driving of a qubit without population inversion.

\section{Non-adiabatic witness in a Quantum Otto cycle}\label{Non-adiabatic witness in a Quantum Otto cycle}

We now move to consider the non-adiabaticity quantity $\mathcal{C}_{\text{diab}}$, based on the quantum texture and imaginarity, in a quantum Otto cycle fueled by a single-qubit as working substance (WS). To establish the notation, let us denote by $\rho(0)$, $\rho(\tau_1)$, $\rho(\tau_2)$, and $\rho(\tau_3)$ the key states of the quantum Otto cycle, i.e., where the quantum texture and the imaginarity are computed. A quantum Otto cycle can be experimentally implemented, for instance, in nuclear magnetic resonance \cite{PhysRevLett.123.240601}, and the coherences in the key states can be accessed using dynamic Bayesian network techniques \cite{PhysRevLett.127.180603}. In this general picture, $\rho(0)$, and $\rho(\tau_2)$ represents the state of the WS after complete thermalization with the thermal reservoirs, here described by Markovian heat baths, while the states $\rho(\tau_1)$ and $\rho(\tau_3)$ represent the WS state after unitary dynamics dictated by a time-dependent Hamiltonian $H(t)$. We are assuming the same during time for the two driving unitary dynamics, such that $\tau_1 - 0 = \tau_3 - \tau_2$.

The non-adiabaticity witness for the entire cycle is the sum for each key states, i.e.,
\begin{equation}
\mathcal{C}^{\text{cycle}}_{\text{diab}} \equiv \sum_{i=0}^3 \left[ \left(2e^{-\mathcal{R}(\tau_i)} - 1\right)^2 + M^2(\tau_i) \right].
\label{witnesscycle}
\end{equation}

Before discussing the finite-time regime, let us first explore the quantity $\mathcal{C}_{\text{diab}}(t)$ for the four key states for a quasi-static quantum Otto cycle. In this paradigmatic case, the WS state remains diagonal in the instantaneous eigenstates of the Hamiltonian throughout the entire cycle, $\rho^{\text{qs}}(\tau_i) = \sum_n \rho_n^{\tau_i} |E_n^{\tau_i}\rangle \langle E^{\tau_i}_n|$. Thus, one has
\begin{eqnarray}
    \sum_{i=0}^3 \mathcal{R}\left[\rho^{\text{ad}}(\tau_i)\right] &=& 4\ln2,\nonumber\\ 
     \sum_{i=0}^3 M\left[\rho^{\text{ad}}(\tau_i)\right] &=& 0,
\end{eqnarray}
leading to $\sum_{i=0}^3 \mathcal{C}_{\text{diab}}(\tau_i) = 0$. In terms of efficiency (or coefficient of performance), this is the best scenario that can be achieved.

The scenario changes considerably in the finite-time regime, where the key states $\rho(\tau_i)$, with $i=1,3$ are no longer diagonal in the instantaneous basis of the driving Hamiltonian $\lbrace {|E_n^{\tau_1}}\rangle\rbrace$ and $\lbrace {|E_n^{\tau_3}}\rangle\rbrace$
, respectively. In this case, these two key states will have coherence in the instantaneous energy basis, which will be well captured by having $\sum_{i=0}^3  \mathcal{C}_{\text{diab}}(\tau_i) > 0$. Note that our definition of non-adiabatic witness prevents a miss interpretation of the scenario where equal amount of coherence is produced in the two unitary strokes but with opposite phase, i.e., $x(\tau_1) = - x(\tau_3)$.

\subsection*{The quantum Otto refrigerator}

\begin{figure}[H]
\includegraphics[width=1.0\linewidth]{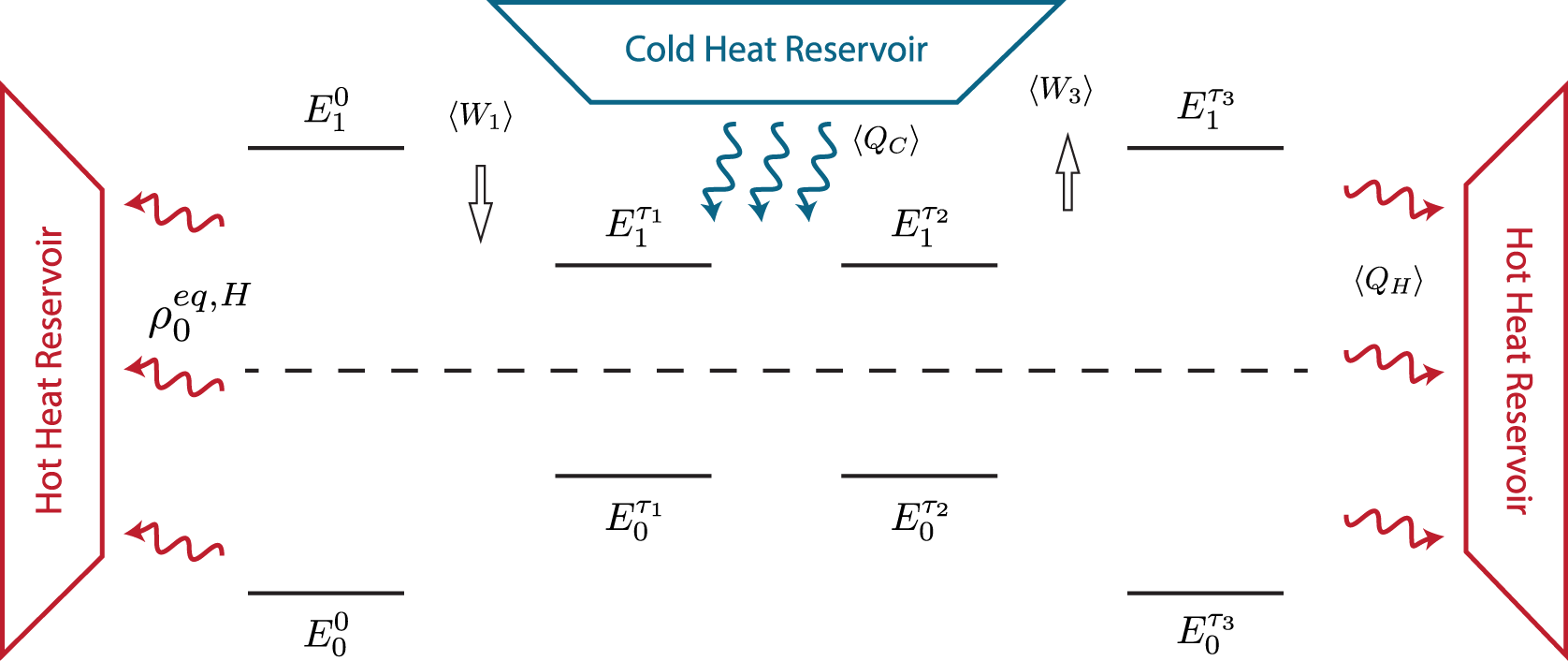}  
\caption{Illustration quantum Otto refrigerator fueled by as single qubit. Initially the working substance is in thermal equilibrium with the hot thermal reservoir. A time-dependent Hamiltonian drives the system up to a frequency $\omega_{\tau_1}$, where it is attached to a cold thermal reservoir. Then, a similar time-dependent Hamiltonian drives back the working substance to its original frequency $\omega_0$, finalizing the cycle with a complete thermalization with the hot thermal bath.}
\label{Scheme}
\end{figure}

The preceding description of a quantum Otto cycle fueled by a single qubit is general and applies to any operation regime. Here we consider in detail a quantum Otto cycle operating as a refrigerator in order to illustrate the use of the proposed non-adiabaticity witness.  The cycle is comprised by two finite-time driven protocols (expansion and compression of the energy gap) and two undriven thermalization process, as illustrated in Fig. \ref{Scheme}. Due to the clear separation of the stages in the Otto cycle, work (heat) is exclusively associated with the driven (undriven) strokes. For the refrigerator cycle, the working substance begins in the hot Gibbs state $\rho\left(0\right) = e^{-\beta_h\, H\left(0\right)}/Z_0^h$, where $\beta_h = \left(k_B T_h\right)^{-1}$ is the hot inverse temperature, $H\left(0\right) = \left(\hbar\omega_0/2\right)\sigma_x$ is the initial Hamiltonian, $\omega_0$ is the initial transition frequency, $\sigma_{x,y,z}$ denotes the Pauli matrices, and $Z_0^h$ is the associated partition function. The complete Otto cycle is described as follows.

\textit{First stroke}. The energy gap of the WS is decreased by the driven Hamiltonian $H^{\text{com}}\left(t\right)$ in a unitary process. To allow the introduction of quantum coherence in the energy basis of the Hamiltonian, we assume the driven Hamiltonian to be given by
\begin{equation}
    H^{\text{com}}\left(t\right) = -\frac{\hbar\omega\left(t\right)}{2}\left[\cos\left(\frac{\pi t}{2\tau_1}\right) \sigma_x + \sin \left(\frac{\pi t}{2\tau_1}\right)\sigma_y \right],
    \label{des01}
\end{equation}
where the time-dependent frequency is modulated by
\begin{equation}
    \omega\left(t\right) = \omega_0 \left(1 - t/\tau_1\right) + \omega_{\tau_1}\left(t/\tau_1\right),
\end{equation}
where the choice of this structure for the driven Hamiltonian and time-dependent frequency is supported by experimental implementations in nuclear magnetic resonance setups \cite{PhysRevLett.123.240601, Denzler2024-qy, Vieira2023-oz}. In this stroke, one could consider that the driven time is fast enough (when compared with the thermalization times) so that the energy exchanged between the WS and the environment can be neglected and hence the driven dynamics is assumed to be unitary. In other words, the system is detached from the thermal reservoirs. The state of the WS after the compression stroke is $\rho\left(\tau_1\right) = U_{\tau_1,0}\rho\left(0\right)U_{\tau_1,0}^\dagger $, with $U_{\tau_1,0} = \mathcal{T}_> \exp\left[-\left(i/\hbar\right)\int_0^{\tau_1}dt\, H^{\text{com}}\left(t\right) \right]$ the time-evolution operator, and $\mathcal{T}_>$ the time-ordering operator. The work associated with this stroke is given by $\langle W_1\rangle = \text{Tr}\left[ H^{\text{com}}\left(\tau_1\right)\rho\left(\tau_1\right)\right] - \text{Tr}\left[H\left(0\right) \rho\left(0\right)\right]$.

\textit{Second stroke}. The WS is attached to the cold thermal reservoir with inverse temperature $\beta_c = \left(k_B T_c\right)^{-1}$ and it undergoes a cold thermalization process. In this stage, the Hamiltonian is kept fixed at $H^{\text{com}}\left(\tau_1\right) = \left(\hbar \omega_{\tau_1}/2\right) \sigma_y$ spanning the time interval $t \in \left[\tau_1,\tau_2\right]$. The cold bath is modeled as a Markovian heat reservoir and the state after the undriven process is given by the master equation \cite{Prathik_Cherian2019-hg}
\begin{eqnarray}
    \frac{d}{dt}\rho\left(t\right) &=& -\frac{i}{\hbar}\left[H\left(t\right),\rho\left(t\right)\right]\nonumber\\
    &+& \gamma_\downarrow\left[\Gamma_\downarrow\rho\left(t\right) \Gamma_\uparrow^\dagger  - \frac{1}{2}\lbrace\rho\left(t\right),\Gamma_\uparrow^\dagger \Gamma_\downarrow \rbrace \right]\nonumber\\
    &+& \gamma_\uparrow\left[\Gamma_\uparrow\rho\left(t\right) \Gamma_\downarrow^\dagger  - \frac{1}{2}\lbrace\rho\left(t\right),\Gamma_\downarrow^\dagger \Gamma_\uparrow \rbrace \right],
    \label{ME}
\end{eqnarray}
with $\gamma_\downarrow = \gamma_0\left(N_{\text{BE}} + 1\right)$, $\gamma_\uparrow = \gamma_0 N_{\text{BE}}$, $\gamma_0$ is the vacuum decay rate, $N_{\text{BE}} = \left(e^{\beta\hbar \omega} -1 \right)^{-1}$ is the Bose-Einstein distribution, and $\Gamma_{\downarrow,\uparrow}$ is the ladder operators in the energy basis. It is possible to analytically solve this master equation and describe the thermalization process through the decomposition on the Bloch sphere. Besides, one assumed complete thermalization with the cold reservoir. In both cases, the heat absorbed by the WS is given by $\langle Q_c\rangle = \text{Tr}\left[H^{\text{com}}\left(\tau_2\right)\rho\left(\tau_2\right) \right] - \text{Tr}\left[H^{\text{com}}\left(\tau_1\right)\rho\left(\tau_1\right) \right]$.

\textit{Third stroke}. The WS energy gap is increased to its original value through a unitary dynamics. The expansion Hamiltonian $H^{\text{exp}}\left(t\right) = H^{\text{com}}\left(\tau_1 + \tau_2 - t\right)$ drives the reverse unitary process. Here, after the driven expansion the WS state is given by $\rho\left(\tau_3\right) = V_{\tau_3,\tau_2}\rho\left(\tau_2\right)V_{\tau_3,\tau_2}^\dagger $, with $V_{\tau_3,\tau_2} = \mathcal{T}_> \exp\left[-\left(i/\hbar\right)\int_{\tau_2}^{\tau_3}dt\, H^{\text{exp}}\left(t\right) \right]$. The work associated with this stroke is $\langle W_3\rangle = \text{Tr}\left[ H^{\text{exp}}\left(\tau_3\right)\rho\left(\tau_3\right)\right] - \text{Tr}\left[H^{\text{exp}}\left(\tau_2\right) \rho\left(\tau_2\right)\right]$.

\textit{Fourth stroke}. Finally, the last process is a complete undriven thermalization with the hot heat reservoir, in order to reset the WS state such that $\rho\left(\tau_4\right) = \rho_0$. Physically, the WS undergoes a similar process described by the master equation in eq. (\ref{ME}), but with the equilibrium thermal state associated with the hot temperature. Here, the heat released to the hot thermal reservoir is $\langle Q_h\rangle = \text{Tr}\left[H^{\text{exp}}\left(\tau_4\right)\rho\left(\tau_4\right) \right] - \text{Tr}\left[H^{\text{exp}}\left(\tau_3\right)\rho\left(\tau_3\right) \right]$.

In the cycle functioning, we adopt the convention that energy flowing in (out) the WS is positive (negative), such that for the cycle operates as a refrigerator, the energetic quantities must satisfy $\langle W_{\text{net}}\rangle = \langle W_1\rangle + \langle W_3\rangle >0$, $\langle Q_c\rangle >0$, and $\langle Q_h \rangle <0$. From an energetic point of view, the performance of a quantum Otto cycle operating in the refrigerator regime is dictated by the coefficient of performance (COP) $\epsilon = \langle Q_c\rangle/\langle W_{\text{net}}\rangle$, the injected power $\langle \mathcal{P}_{\text{injected}}\rangle = \langle W_{\text{net}}\rangle / \tau_{\text{cycle}}$, and the cooling rate $\langle \mathcal{P}_{\text{cooling}}\rangle = \langle Q_c\rangle / \tau_{\text{cycle}}$, with $\tau_{\text{cycle}}$ the total time of the cycle. In the present model, we are assuming complete thermalization in the undriven strokes, such that the relevant time scale to compute the the powers is the time $\tau_1$.

In addition to the energetic quantities, the non-commutative aspect of the driven Hamiltonian in different times, along with the finite-time dynamics, will be responsible for the generation of coherence in the energy basis in the first and third strokes. By considering complete thermalization with the cold thermal reservoir, only the key states $\rho\left(\tau_1\right)$ and $\rho\left(\tau_3\right)$ have coherence in the instantaneous energy basis of the Hamiltonian. 

    The COP can be related to the non-adiabaticity witness at the end of the first and third strokes, $\mathcal{C}_{\text{diab}}(\tau_1)$ and $\mathcal{C}_{\text{diab}}(\tau_3)$, respectively.  In the Appendix \ref{B} we derive the following relation
    \begin{equation}
    \frac{1}{\epsilon} =\frac{1}{\epsilon_{\text{Otto}}} - \mathcal{L}_{qs},
    \label{res04}
\end{equation}
where $\mathcal{L}_{qs}$ represents the COP lag, quantifying how disparate is the COP from the quasi-static COP, $\epsilon_{\text{Otto}} =  \omega_{\tau_1} /(\omega_0 - \omega_{\tau_1})$, and is given by
\begin{eqnarray}
    \mathcal{L}_{qs} &\equiv& \frac{\hbar\omega_{0}}{2\langle Q_c\rangle} \frac{\mathcal{C}_{\text{diab}}(\tau_1)}{\Delta p_h + \sqrt{\Delta p_h^2 - \mathcal{C}_{\text{diab}}(\tau_1)}}\nonumber\\
    &+&  \frac{\hbar\omega_{0}}{2\langle Q_c\rangle}\frac{\mathcal{C}_{\text{diab}}(\tau_3)}{\Delta p_c + \sqrt{\Delta p_c^2 - \mathcal{C}_{\text{diab}}(\tau_3)}},
    \label{res05}
\end{eqnarray}
where $\Delta p_c = \tanh \left(\beta_c \hbar \omega_{\tau_1}/2\right)$ and  $\Delta p_h = \tanh \left(\beta_h \hbar \omega_{0}/2\right)$.

Moreover, using the definitions of injected power and cooling rate we obtain the following expressions in terms of the Otto lag
\begin{eqnarray}
    \langle \mathcal{P}_{\text{injected}}\rangle &=& \frac{\langle Q_{c}\rangle}{\tau_{cycle}}\left(\frac{1}{\epsilon_{Otto}}-\mathcal{L}_{qs}\right),\nonumber\\
    \langle \mathcal{P}_{\text{cooling}}\rangle &=& \frac{\epsilon_{Otto}}{\left(1-\epsilon_{Otto}\mathcal{L}_{qs}\right)}\frac{\langle W_{net}\rangle }{\tau_{cycle}}.
\end{eqnarray}

Equations (\ref{res04}) and (\ref{res05}) together show explicitly the role played by the quantum texture and imaginarity in the performance of the quantum Otto refrigerator fueled by a single qubit. The coherences produced along the unitary strokes are distributed  into quantum texture and imaginarity of the WS working substance.

\begin{figure}[h!]
    \centering
    \includegraphics[width=0.45\textwidth]{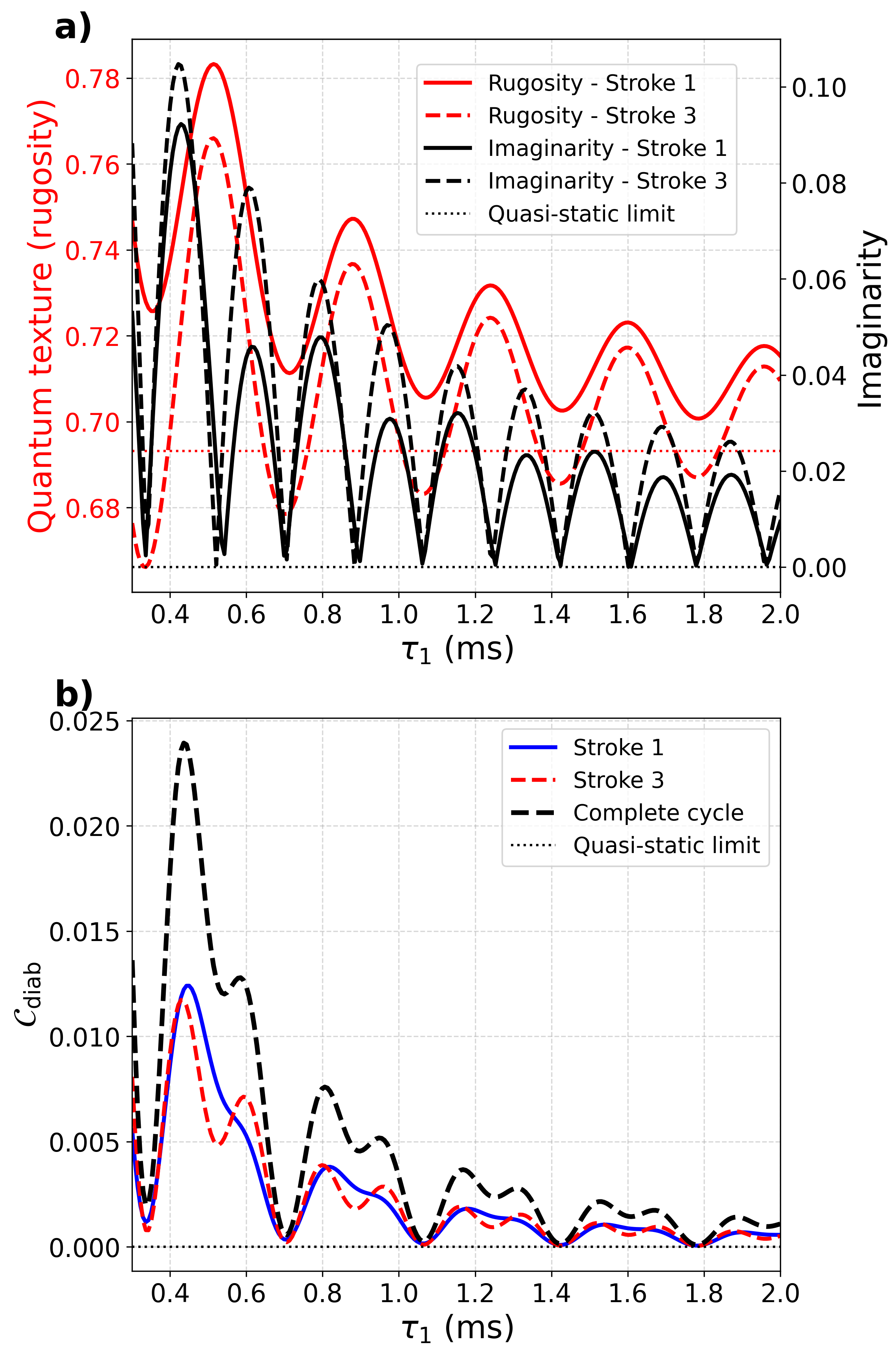}
    \caption{Dynamics of non-adiabatic quantities along the quantum Otto refrigerator as a function of $\tau_1$. a) Quantum texture (rugosity) $\mathcal{R}$ and imaginarity $M$. b) Non-adiabaticity witness $\mathcal{C}_{\text{diab}}$. In both figures, we computed quantities for the key states $\rho(\tau_1)$, $\rho(\tau_3)$. The quantity $\mathcal{C}_{\text{diab}}$ is computed also for the entire quantum Otto cycle. We have used the following parameters: $\omega_0 = 2.0$kHz, $\omega_{\tau_1} = 3.6$kHz, $T_c = 5\hbar\omega_{\tau_1}/(2k_B)$, and $T_h = 2\hbar\omega_{0}/(2k_B)$.}
    \label{quantities}
\end{figure}

Figure \ref{quantities}-\textbf{a) } depicts the quantum texture (rugosity) (red lines) and imaginarity (black lines), for the key states $\rho\left(\tau_1\right)$ (solid) and $\rho\left(\tau_3\right)$ (dashed) at the end of the unitary strokes as a function of $\tau_1$. As expected, for long driving times, the quantum texture for both key states approaches $\ln 2$, while the imaginarity vanishes. Furthermore, Fig.  \ref{quantities}-\textbf{b)} shows the proposed non-adiabaticity witness based on the quantum texture and imaginarity. For every driving time, it remains positive in the three cases, approaching zero for long driving times, when the cycle becomes quasi-static. Importantly, note that the individual behavior of the quantum texture or imaginarity fails to fully capture the non-adiabaticity for a given time $\tau_1$. 

\begin{figure}[h!]
    \centering
    \includegraphics[width=0.45\textwidth]{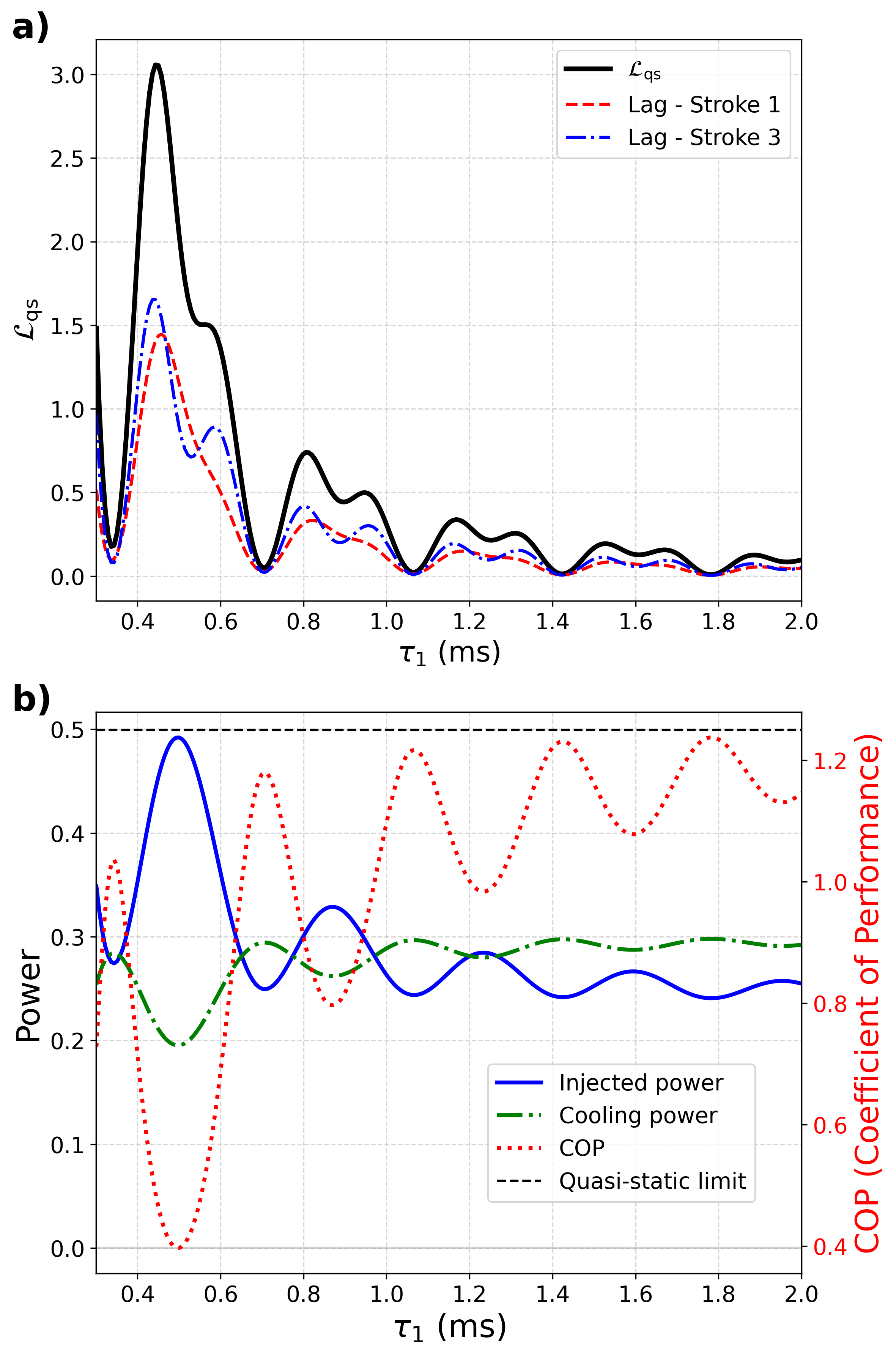}
    \caption{a) Otto Lag as Eq. (\ref{res05}). The total Otto lag (solid black line) and its components relative to the first stroke (dashed red line) and third strokes (dashed-dot blue line). b) Performance of the quantum Otto refrigerator. The COP (dotted red line), the injected power (solid blue line), and the cooling rate (dashed-dot green line). In both pannels the quantities are plotted as function of $\tau_1$. The parameters of the cycle are the same as in Fig. \ref{quantities}. }
    \label{energetics}
\end{figure}

Extending the analysis, Fig. \ref{energetics}-\textbf{a)} depicts the total Otto Lag (solid black line) and its components relative to the first stroke (dashed red line) and third strokes (dashed-dot blue line) as function of $\tau_1$, whereas in Fig. \ref{energetics}-\textbf{b)} we illustrate the performance of the quantum Otto refrigerator by showing the coefficient of performance (dotted red line), the injected power (solid blue line) and the cooling rate (dashed-dotted green line), all as function of $\tau_1$. We also indicate the quasi-static COP (dashed black line). Whereas the COP deviates considerably from the quasi-static limit for short driven times, it approaches this limit for long driven times. The oscillations in the three quantities represents the generation of coherence along the unitary strokes, in the form of quantum texture and imaginarity. It must be stressed that the different local maxima of the COP in Fig. \ref{energetics}-\textbf{b)} correspond numerically to the different local minima in  Fig.\ref{quantities}-\textbf{b)} for the quantity $\mathcal{C}^{\text{cycle}}_{\text{diab}}$, corroborating the use of this quantity also to indicate better (or worse) operating regimes.

\section{Conclusions}\label{Conclusion}

In this work, we have analyzed the role played by the quantum texture (rugosity) and imaginarity in the performance of a quantum Otto cycle fueled by a single qubit as working substance. By considering finite-time driven operations and complete thermalization with the heat sources, we have derived analytical expressions relating the performance of the quantum Otto cycle to the quantum friction written in terms of the quantum texture and imaginarity of the WS quantum state along the cycle. Although we detailed the expressions for the refrigerator regime, our results are valid for any operation regime or even a single protocol, provided the system is a qubit.

An important aspect is that quantum texture or imaginarity alone fails to correctly capture the inner friction of the working substance. Because of that, we proposed a non-adiabaticity witness to jointly indicate the quantum coherence along a process or an entire cycle, associating quantum texture and imaginarity. The quantity proposed here fulfills important properties, such that it is zero for quasi-static process, strictly non-negative, and monotonically under incoherent CPTP maps. Furthermore, it is experimentally accessible through quantum state tomography in platforms such as nuclear magnetic resonance.

Interesting directions may be explored in the future. An appealing avenue for future research is the generalization of our framework beyond single-qubit working substances to multi-level quantum systems ($d>2$). While our derivation relies on the Bloch sphere representation, both the operational resource theory of imaginarity and the quantum texture admit formulations in arbitrary $d$-dimensional Hilbert spaces. In addition, the generalization for quantum Otto cycles with incomplete thermalization with the heat sources could be useful, exploring quantum lubrication techniques . Finally, exploring shortcuts to adiabaticity (STA) — such as counterdiabatic driving protocols designed to cancel $\mathcal{C}_{\text{diab}}$ at short operation times — could offer practical pathways for reaching the ideal Otto performance without sacrificing cooling power.

We hope that our work can contribute to unveil the role played by quantum texture and imaginarity in quantum processes as well as their connection with thermodynamics and irreversibility.

\begin{acknowledgments}
Jonas F. G. Santos acknowledges CNPq Grant No. 420549/2023-4, Fundect
Grant No. 83/026.973/2024, and Universidade Federal da Grande Dourados
for support.
\end{acknowledgments}

\FloatBarrier
\appendix
\section{Derivation of the relations (\ref{res01}) and (\ref{res02})}
\label{A}

Consider a qubit in which unitary dynamics is given by the von Neumann equation
\begin{equation}
    i\hbar \frac{d}{dt}\rho(t) = \left[H(t),\rho(t)\right].
\end{equation}

At $t = 0$ the initial state is written as 
\begin{equation}
    \rho(0) = \frac{1 + \Delta p_0}{2}|E_0^0\rangle \langle E_0^0| + \frac{1 - \Delta p_0}{2}|E_1^0\rangle \langle E_1^0|,
\end{equation}
with $\Delta p_0 = p_0(0) - p_1(0)$. At time $t = \tau$, the instantaneous basis is $\lbrace{|E_0^\tau\rangle, |E_1^\tau\rangle }\rbrace$. The Hamiltonian in this basis is
\begin{equation}
    H(\tau) = E^\tau \mathbb{I} - \frac{\Delta^\tau}{2}\sigma_z^\tau,
\end{equation}
with $E^\tau = (E_0^\tau + E_1^\tau)/2$, $\Delta^\tau = E_1^\tau - E_0^\tau >0$, and $\sigma_z^\tau = |E_0^\tau\rangle \langle E_0^\tau| - |E_1^\tau\rangle \langle E_1^\tau|$.

The qubit state at time $t = \tau$ in the instantaneous basis is
\begin{eqnarray}
    \rho(\tau) &=& \sum_{n,m=0}^1 \rho_{nm}(\tau) |E_n^\tau\rangle \langle E_m^\tau|\nonumber\\
    &=&\frac{1}{2}\left[\mathbb{I}+ r_x(\tau) \sigma_x^\tau + r_y(\tau) \sigma_y^\tau + r_z(\tau) \sigma_z^\tau\right],
\end{eqnarray}
with the elements in the Bloch sphere
\begin{eqnarray}
    r_x(\tau) &=& 2\text{Re}\left[\rho_{01}(\tau)\right],\nonumber\\
    r_y(\tau) &=& -2\text{Im}\left[\rho_{01}(\tau)\right],\nonumber\\
    r_z(\tau) &=& \rho_{00}(\tau) - \rho_{11}(\tau).\nonumber\\
\end{eqnarray}

On the other hand, if the unitary dynamics is slow enough to be quasi-static, the state keeps diagonal in the instantaneous basis without changing its population. In this case, the state is given by
\begin{equation}
    \rho_{\text{qs}}(\tau) = \frac{1 + \Delta p_0}{2}|E_0^\tau\rangle \langle E_0^\tau| + \frac{1 - \Delta p_0}{2}|E_1^\tau\rangle \langle E_1^\tau|.
\end{equation}

The work associated with the inner friction is 
\begin{eqnarray}
    \langle W_{\text{fric}}(\tau)\rangle  &=& \langle W (\tau)\rangle  - \langle W(\tau)\rangle_{\text{qs}}\nonumber\\
    &=& \langle H(\tau) \rangle - \langle H(\tau) \rangle_{\text{qs}}
\end{eqnarray}
and computing the averages,
\begin{eqnarray}
    \langle H(\tau) \rangle &=& \text{Tr}\left[\rho(\tau) H(\tau)\right] = E^\tau - \frac{r_z(\tau)}{2}\Delta^\tau,\nonumber\\
    \langle H(\tau) \rangle_{\text{qs}} &=& \text{Tr}\left[\rho_{\text{qs}}(\tau) H(\tau)\right] = E^\tau - \frac{\Delta p_0}{2}\Delta^\tau,\nonumber\\
\end{eqnarray}
such that
\begin{equation}
    \langle W_{\text{fric}}(\tau)\rangle = \frac{\Delta^\tau}{2}\left[\Delta p_0 - r_z(\tau)\right].
\end{equation}

Now, note that for any unitary dynamics it follows that $|r(\tau)|^2 = r_x^2(\tau) + r_y^2(\tau) + r_z^2(\tau) = |r(0)|^2 = \Delta p_0^2$, such that one has
\begin{eqnarray}
    \Delta p_0 - r_z(\tau) &=& \frac{r_x^2(\tau) + r_y^2(\tau)}{\Delta p_0 + \sqrt{\Delta p_0^2 - \left[r_x^2(\tau) + r_y^2(\tau)\right]}}\nonumber\\
    &=& \frac{\mathcal{C}_{\text{diab}}(\tau)}{\Delta p_0 + \sqrt{\Delta p_0^2 - \mathcal{C}_{\text{diab}}(\tau)}},
\end{eqnarray}
where the last line follows from Eqs. (\ref{cc}) and (\ref{ccc}). Hence, the work associated with the inner friction reads
\begin{equation}
    \langle W_{\text{fric}}\rangle = \frac{\Delta^\tau}{2} \frac{\mathcal{C}_{\text{diab}}(\tau)}{\Delta p_0 + \sqrt{\Delta p_0^2 - \mathcal{C}_{\text{diab}}(\tau)}}.
    \label{A00}
\end{equation}

By assuming $E_0^\tau = -E_1^\tau = -\hbar\omega(\tau)/2$ and that the qubit is initially in a thermal state with inverse temperature $\beta_0 = (k_B T_0)^{-1}$, we immediately get the Eq. (\ref{res01}) in the main text.

Next, by noting that $\Delta p_0^2 - r_z^2(\tau) = \left[\Delta p_0 + r_z(\tau)\right]\left[\Delta p_0 - r_z(\tau)\right] \geq \left[\Delta p_0 - r_z(\tau)\right]^2$, it follows that $\Delta p_0 - r_z(\tau) \leq \sqrt{\Delta p_0^2 - r_z^2(\tau)}$, provided $r_z(\tau) \geq 0$ (no population inversion). Then, using the preceding results, we obtain Eq. (\ref{res02}) in the main text
\begin{equation}
    \langle W_{\text{fric}}(\tau)\rangle \leq \frac{\hbar \omega(\tau)}{2} \sqrt{\mathcal{C}_{\text{diab}}(\tau)}.
\end{equation}

\section{Relation between COP and $\mathcal{C}_{\text{diab}}$ }
\label{B}

First, let us assume a driven protocol from $t = 0$ to $t = \tau$ followed by a complete thermalization with a Markovian heat bath at inverse temperature $\beta$. From the inner friction
\begin{eqnarray}
    \langle W_{\text{fric}}(\tau)\rangle &=& \langle W (\tau)\rangle - \langle W (\tau)\rangle_{qs}\nonumber\\
    &=&\langle H(\tau) \rangle - \langle H(\tau)\rangle_{qs},
\end{eqnarray}
the heat exchanged with the thermal bath can be expressed as 
\begin{eqnarray}
    \langle Q\rangle &=& \langle H(\tau_{\text{therm}})\rangle  - \langle H(\tau) \rangle\nonumber\\
    &=&\langle Q\rangle_{qs} - \langle W_{\text{fric}}(\tau)\rangle,
\end{eqnarray}
with $\langle Q\rangle_{qs} = \langle H(\tau_{\text{therm}})\rangle - \langle H(\tau)\rangle_{qs}$ the exchanged heat if the system state is diagonal in the instantaneous basis at $t = \tau$. Here $\tau_{\text{therm}}$ is a thermalization time larger enough to ensure complete thermalization.

For the quantum Otto cycle as a refrigerator, using this relation for the complete thermalization with the cold thermal bath after the first stroke and for the complete thermalization with the hot thermal bath after the third stroke, we obtain
\begin{eqnarray}
    \langle Q_c \rangle &=& \langle Q_c\rangle_{qs} - \langle W_{\text{fric}}^1\rangle,\nonumber\\
    \langle Q_h \rangle &=& \langle Q_h\rangle_{qs} - \langle W_{\text{fric}}^3\rangle.
    \label{B01}
\end{eqnarray}

From the definition of the coefficient of performance (COP) for the refrigerator,
\begin{eqnarray}
    \epsilon = \frac{\langle Q_c\rangle}{\langle W_{\text{net}}\rangle},
\end{eqnarray}
and using the first law of thermodynamics, we can write the inverse of the COP as
\begin{equation}
    \frac{1}{\epsilon} =  \left(1 + \frac{\langle Q_h\rangle}{\langle Q_c\rangle}\right).
    \label{B02}
\end{equation}

During the thermalization strokes, the energy gap is kept fixed. Besides, a quasi-static quantum Otto cycle has the property  $\langle Q_h\rangle _{qs}/\langle Q_c\rangle _{qs} = -\omega_0/\omega_{\tau_1}$, and employing (\ref{B01}) we get
\begin{equation}
    \frac{\langle Q_h\rangle}{\langle Q_c\rangle} = -\frac{\omega_0}{\omega_{\tau_1}} - \frac{1}{\langle Q_c\rangle} \left[\frac{\omega_0}{\omega_{\tau_1}}\langle W_{\text{fric}}^1\rangle + \langle W_{\text{fric}}^3\rangle \right],
\end{equation}
and substituting this expression into (\ref{B02}),
\begin{equation}
    \frac{1}{\epsilon} =\frac{1}{\epsilon_{\text{Otto}}} - \mathcal{L}_{qs},
\end{equation}
with $\epsilon_{\text{Otto}} =  \omega_{\tau_1} /(\omega_0 - \omega_{\tau_1})$ the corresponding COP for the quasi-static quantum Otto refrigerator and
\begin{equation}
    \mathcal{L}_{qs} \equiv \frac{1}{\langle Q_c\rangle} \left[\frac{\omega_0}{\omega_{\tau_1}}\langle W_{\text{fric}}^1\rangle + \langle W_{\text{fric}}^3\rangle \right]
    \label{B03}
\end{equation}
the COP lag, quantifying how disparate the COP is from its maximum value.

Finally, from (\ref{A00}) we immediately relate the COP lag to the non-adiabaticity quantity in terms of the quantum texture and imaginarity,
\begin{eqnarray}
    \langle W_{\text{fric}}^1\rangle &=& \frac{\hbar\omega_{\tau_1}}{2} \frac{\mathcal{C}_{\text{diab}}(\tau_1)}{\Delta p_h + \sqrt{\Delta p_h^2 - \mathcal{C}_{\text{diab}}(\tau_1)}},\nonumber\\
    \langle W_{\text{fric}}^3\rangle &=& \frac{\hbar\omega_{0}}{2} \frac{\mathcal{C}_{\text{diab}}(\tau_3)}{\Delta p_c + \sqrt{\Delta p_c^2 - \mathcal{C}_{\text{diab}}(\tau_3)}},
\end{eqnarray}
i.e., the equations presented in the main text.

\bibliographystyle{apsrev4-2}
\bibliography{refs}

\end{document}